# The risk ethics of autonomous vehicles: a continuous trolley problem in regular road traffic

Sebastian Krügel

Faculty of Computer Science, Technische Hochschule Ingolstadt, Germany School of Governance, Technical University of Munich, Germany

Matthias Uhl

Faculty of Computer Science, Technische Hochschule Ingolstadt, Germany

May 24, 2022

## Abstract

Is the ethics of autonomous vehicles (AVs) restricted to weighing lives in unavoidable accidents? We argue that AVs distribute risks between road users in regular traffic situations, either explicitly or implicitly. This distribution of risks raises ethically relevant questions that cannot be evaded by simple heuristics such as "hitting the brakes." Using an interactive, graphical representation of different traffic situations, we measured participants' preferences on driving maneuvers of AVs in a representative survey in Germany. Our participants' preferences deviated significantly from mere collision avoidance. Interestingly, our participants were willing to take risks themselves for the benefit of other road users suggesting that the social dilemma of AVs may lessen in a context of risk.

**Corresponding author:** Sebastian Krügel, Faculty of Computer Science, Technische Hochschule Ingolstadt, Esplanade 10, 85049 Ingolstadt, Germany, Email: sebastian.kruegel@thi.de

# Introduction

Deterministic dilemmas resembling the trolley problem (Foot, 1967) still predominate ethical discussions on AVs. This has been criticized as it neglects that road traffic is not deterministic, but risky (Goodall, 2016a; Nyholm & Smids, 2016; Trussel, 2018; Keeling et al., 2019; Geisslinger et al., 2021). Connected AVs may be considered a chance for managing traffic risk more deliberately than impulse-driven manual traffic allows. By focusing on unavoidable accidents, however, the recent ethical perspective sets accident probability to one and solely discusses accident severity. This may be understandable, because risk was traditionally located in the realm of rationality and attributed to decision theory instead of ethics that focused on actual harm (MacLean, 2012; Hansson, 2013, 2018). At the other extreme, the engineering perspective is essentially guided by accident avoidance (Reichardt & Shick, 1994; Funke et al., 2015; Gerdes & Thornton, 2015; Thornton et al., 2017) and therefore questions the relevance of the ethics of unavoidable accidents altogether (Trussel, 2018; Winfield et al., 2019). The focus of engineering is on minimizing accident probability. Both perspectives neglect that any maneuver in regular road traffic constitutes a redistribution of risks which is a function of both accident probability and accident severity.

To incorporate moral folk intuitions into the ethical debate, laypeople's choices in trolley problems with AVs have been elicited (Bonnefon, Shariff & Rahwan, 2016; Awad et al., 2018; Bigman & Gray, 2020; Krügel, Uhl & Balcombe, 2021; Krügel & Uhl, 2022). These results can be transferred to regular traffic situations if one interprets them as statistical trolleys (Bonnefon et al., 2019). If a certain traffic situation occurs million times, accidents will be unavoidable in the aggregate. The aim of the present paper is to investigate individual choices under risk explicitly by letting people take moral decisions in regular road traffic. By looking at safety distances, we can directly study the rationale with which people manage the trade-offs between accident probability and accident severity without assuming one of the components away. Driving algorithms factually distribute risks in road traffic among different road users when determining their safety distance between them (Goodall, 2016b; Bonnefon et al., 2019; Geisslinger et al., 2021; Krügel & Uhl, 2022). It seems more sensible to reflect on the principles underlying this distribution of risk through an ethical discourse than to leave it to the explicit or implicit judgments of car manufacturers (Dolgov & Urmson, 2014).

For the purpose of contributing to this ethical discourse, we developed a graphical interface depicting a common traffic situation in a possible future with AVs operating in mixed traffic. Using this interface, we elicited laypeople's intuitions about the distribution of traffic risks. *Figure 1* shows examples of four traffic situations out of a total of 29 that we have used in our study. In all traffic situations, a self-driving (yellow) car was depicted between two other road users and its driving position between them could be gradually adjusted by the participants in 99 increments by dragging the yellow car to the left and right or using the arrow keys below it. Participants were told that the overall probability of an accident was very small, but not zero, and that the probability of a collision with either vehicle depended on the driving distance of the yellow car. The smaller this distance, the greater the probability of a collision with that vehicle, assuming that the yellow car cannot collide with both vehicles at the same time. Furthermore, it should be presumed that in the event of a collision, all parties involved in the accident are dead.

We presented the task of positioning the AV to a representative sample in Germany.[1] We checked the participants' proper understanding of the underlying situation with the help of two control questions. Only participants who answered both questions correctly were finally able to complete the survey. Out of a total of 4,104 participants, 1,807 participants answered both control questions correctly and became part of the study. Each participant assessed only one traffic situation and the initial position of the yellow car was chosen at random for each participant. We varied the traffic situations along three dimensions. That was, first, the number of passengers in the other two cars to the left and right of the yellow AV. That was second, the type of vehicle on the right side of the road (a car with one passenger or one cyclist). And finally, whether the participants themselves were part of the traffic situation by being a passenger in the yellow AV in some treatments but not in others. Except in the treatments where a cyclist was shown on the right side, we also mirrored each traffic situation so that the majority of passengers were sometimes shown on the left and sometimes on the right side of the road.

---

[1] We obtained approval for the study by the Institutional Review Board of the German Association for Experimental Economic Research e.V. The study was pre-registered on 'https://aspredicted.org.'

To highlight the possible rationale of mere accident avoidance in our traffic situations graphically, we visualized the collision probability with the left or right vehicle using red bars below the respective vehicle. The middle driving position of the AV between the two other vehicles minimized the overall accident probability. This probability grew exponentially with the deviation from the middle driving position. The shorter the distance to one of the two other vehicles, the greater the increase in the probability of a collision with this vehicle.

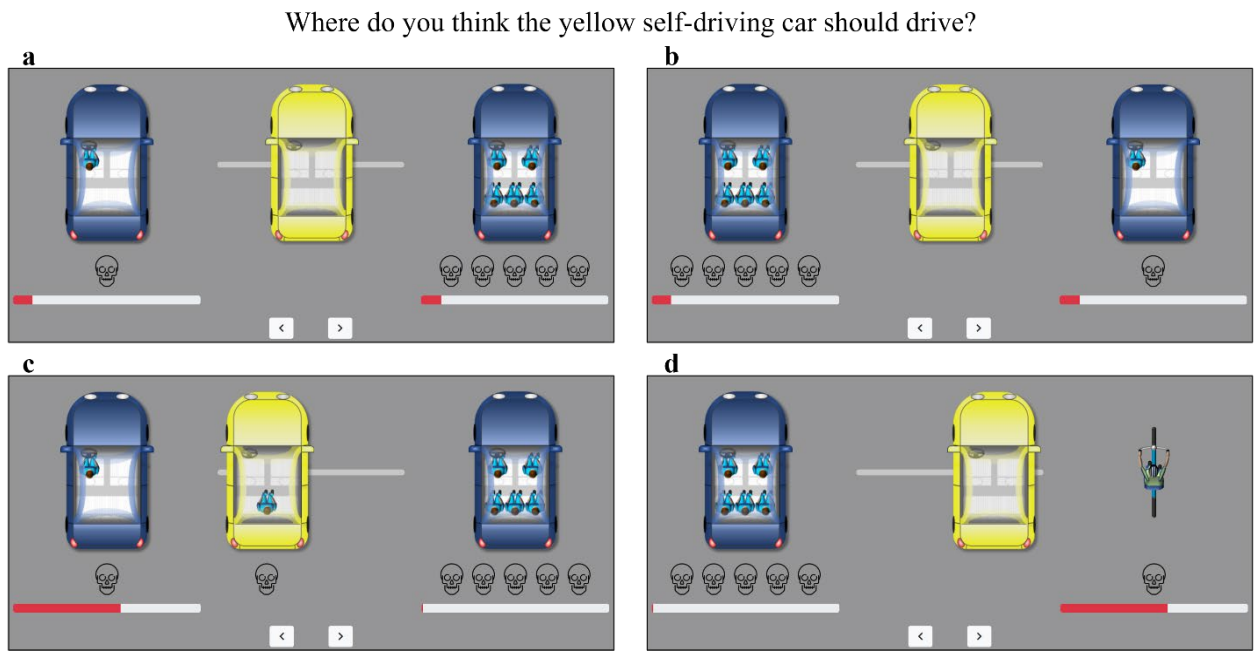

**Fig. 1. Graphical interface for eliciting preferences about risk allocation in road traffic.** The figure shows four possible traffic situations of a self-driving car. The probability of a collision with another road user is very small, but not zero. The following rule applies: The smaller the distance, the higher the probability of a collision. If a collision occurs, all those affected by the collision are dead. The self-driving car cannot collide with both other road users at the same time. By dragging the yellow car back and forth and by using the arrow keys, participants could continuously adjust the driving position between the two other road users.

# Results

*Figure 2a* shows the participants' average driving position of the AV between the two other cars, where the number of passengers in the latter varied. Focusing first on the treatments in which the participants were not part of the traffic situation (see AVs with red frames (= empty AVs)), it is evident that, on average, participants considered the number of passengers in the two other cars in their AV positioning. Comparing each of the pairwise, mirrored traffic situations that differed only in terms of whether the majority of passengers in the cars appeared to be on the left or right side of the road, the average driving position of the AV was significantly different in all of these paired situations. On average, participants positioned the AV always closer to the car with fewer passengers ($t > 3.4$, $p < 0.0011$, for each of the paired situations). While participants placed the AV virtually at the middle driving position when there was one passenger in the car on both sides of the road, the driving position in all other treatments deviated from the equidistant, middle position. Except for the treatments *1 vs. 2* ($t = 1.83$, $p = 0.08$), *1 vs. 3* ($t = 1.79$, $p = 0.08$), and *1 vs. 5* ($t = 0.75$, $p = 0.46$), the deviations from the middle driving position were statistically significant in all other treatments ($t > 3.08$, $p < 0.0037$, for each of the other tests). The red line in *Figure 2b* visualizes a significant positive trend in the safety distance of the AV to the car with more passengers the larger the disproportion of passengers on both sides was ($p < 0.001$).

Surprisingly, the results are very similar when participants were asked to imagine that they themselves were part of the traffic situation (see AVs with blue frames (= AVs with one passenger) in *Figure 2a*). Again, the average driving position of the AV always differed significantly between the mirrored traffic situations and was closer to the car with fewer passengers in each case ($t > 2.49$, $p < 0.015$, for each of the paired situations). In five of the eight treatments with an imbalance in the number of passengers on the left and right sides (*2 vs. 1*, *3 vs. 1*, *4 vs. 1*, *5 vs. 1*, *1 vs. 5*), the AV's preferred driving position was significantly different from the lane's middle ($t > 2.30$, $p < 0.026$, for each of these tests) – the driving position with the lowest probability of an accident. Overall, we see again a significant positive trend between the AV's safety distance to the car with more passengers and the disproportion of passengers in the two other cars in these treatments (see *Figure 2b*). In fact, the two trend lines with 95% confidence bands in *Figure 2b* suggest that there were no significant differences between the treatments where participants were passengers of the AV and where they were neutral observers of the traffic situation. This is surprising because we know from previous studies with deterministic trolley problems that people prefer riding with AVs that protect them as passengers at all costs (Bonnefon, Shariff & Rahwan, 2016).[2]

---

[2] To confirm that the participants understood that the middle driving position minimized the probability of an accident, we re-invited some randomly selected participants from our main study and described the situation to these participants once again. In the follow-up study, we implemented two treatments, both of which were based on the same traffic situation. In each case, the participants were passengers sitting in the (yellow) AV and there were five passengers in the (blue) car on the left and one passenger in the (blue) car on the right side of the road. In one treatment, participants could again adjust the driving position of the AV between the two other cars in 99 increments. This time, however, we did not ask them where they thought the AV should drive, but where the probability of an accident was the lowest for them as a passenger in the (yellow) AV. On average, the participants (n = 51) positioned the AV exactly in the middle between the other two cars (mean = 49.5, s.e. = 1.75). In the other treatment, participants could not position the AV themselves, but they saw two pictures. In one picture, the AV was driving exactly between the other two cars, and in the other picture it was driving on the far right and thus closer to the blue car with one passenger. When asked in which of the two pictures the probability of an accident would be the lowest for the participants as a passenger in the AV, 87.7% (= 50 out of 57) chose the correct picture. 3.5% (= 2 of 57) chose the incorrect picture and 8.8% (5 of 57) indicated that the probability would be the same in both pictures. Overall, the results clearly show that almost all participants had understood where the lowest accident probability for the AV was in our traffic situations.

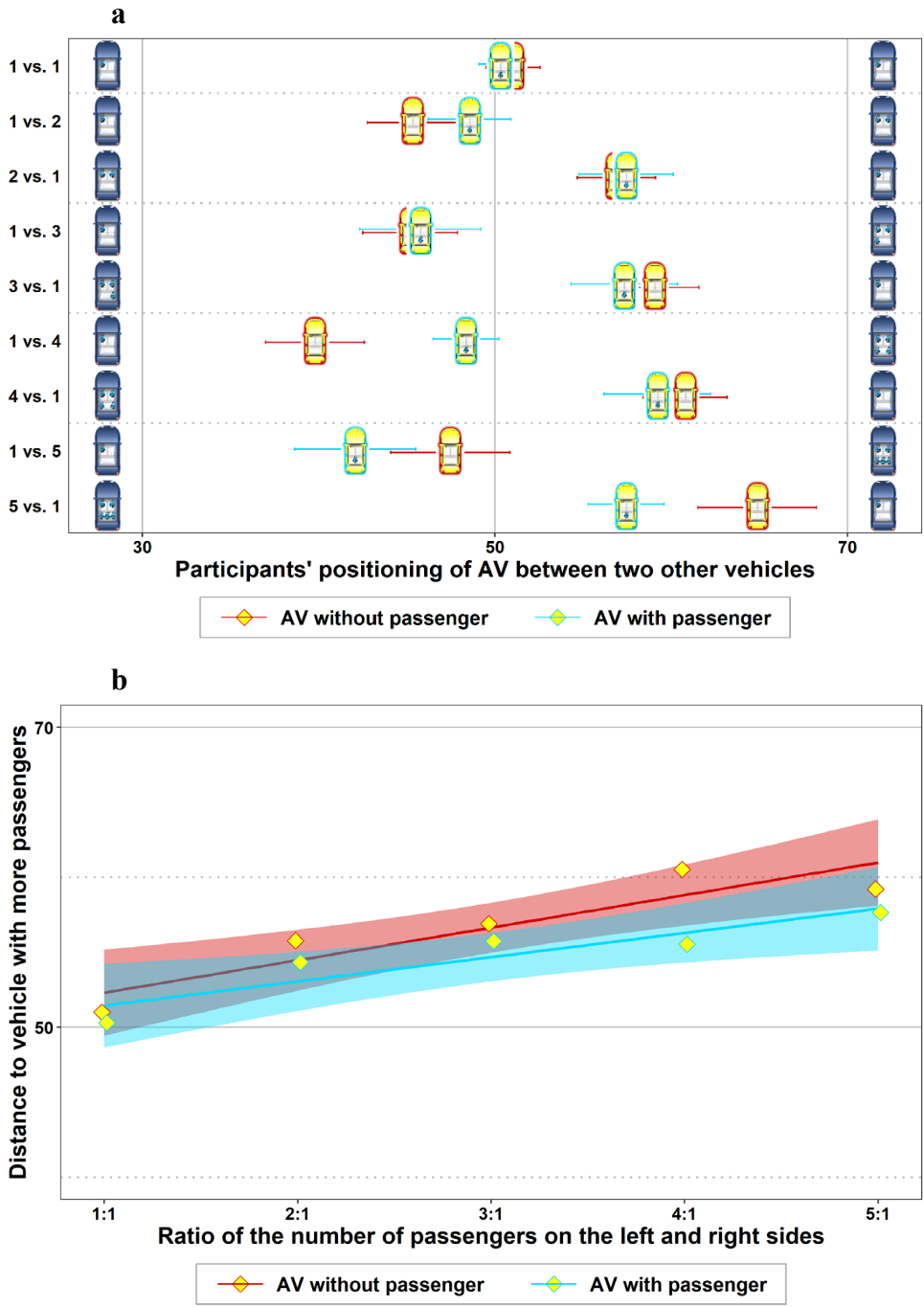

**Fig. 2. Risk allocation between road users of the same type.**
(a) shows the means and standard errors of participants' chosen driving position of the AV in traffic situations with different numbers of passengers in the blue cars on the left and right sides. AVs with red frames depict the results when the AVs were empty. AVs with blue frames depict the results where the participants themselves were a passenger in the AV. (b) shows the results of a linear regression in which the AV's distance from the car with the majority of passengers was regressed on the increasing disproportion of passengers on the left and right sides. We interacted the disproportion of passengers with a dummy variable for the two treatments (passenger in AV yes/no). Red and blue regions visualize 95% confidence intervals.

*Figure 3a* shows participants' average driving positions of the AV for different vehicle types on the right side of the road. AVs with green frames show the average positions when there was a cyclist and AVs with red frames when there was a car with one passenger on the right side. Overall, cyclists received a small bonus in the distribution of risks from our participants. If the AV traveled between a car with one passenger and one cyclist, the latter was granted marginally more safety distance, at the expense of the car passenger (*1 vs. 1*: $t = 1.72$, $p = 0.088$). As the number of car passengers increased on the left side, the AV was positioned closer to the cyclist, but this adjustment was smaller than when there was a car with one passenger on the right side of the road. In Treatments *4 vs.1* and *5 vs. 1*, this resulted in significant different driving positions between the treatments with a cyclist and a car passenger ($t > 2.18$, $p < 0.031$, in both tests). In both treatments, we see overall a significant positive trend in the safety distance of the AV to the car on the left side of the road the more passengers were riding in that car (see *Figure 3b*). However, this positive trend is slightly less pronounced when there was a cyclist on the ride side of the road. Note that the risk bonus of the cyclist cannot easily be attributed to the cyclist's higher vulnerability, because an accident in our scenarios was always fatal for all parties involved and this understanding was verified among our participants. The small risk bonus for cyclists in our study is consistent with the results of Awad et al. (2018), where a weak preference for pedestrians in moral dilemmas with AVs was reported. Normatively, the cyclist's risk bonus could be justified with his or her lower imposition of traffic risks on other road users.

*Table 1* shows the regression results for each treatment (*AV empty*, *AV with passenger*, *bike on the right side*), where the dependent variable is the distance to the vehicle with more passengers.[3] The greater the disproportion of passengers in the two vehicles on the left and right sides of the road, the greater the safety distance to the vehicle with more passengers. In addition, in the *AV empty* and *AV with passenger* treatments, there was some restraint overall in deviating from the lane's middle when there were more passengers on the right, as opposed to the left side of the road. Possibly, this peculiarity stems from an internalization of the emergency lane, which in Germany, as well as in some other European countries, must always be formed between the far left lane and the other lanes. Apart from this restraint, only the "bat and ball problem"[4] of the cognitive reflection test (Frederick, 2005) was significantly associated with AV positioning. Participants who solved this problem correctly incorporated the number of passengers more strongly into their decision and were thus somewhat more utilitarian, which is generally consistent with conjectures in the literature (see, e.g., Greene, 2014). Other demographic variables, the participants' risk attitude,[5] their decision time in positioning the AV, and the randomized initial position of the AV did not affect the final positioning between the two other vehicles.

---

[3] If there were the same number of passengers on the right and left sides, the dependent variable is the distance from the left vehicle. The results remain the same if the distance from the right vehicle is taken instead in these cases.

[4] The "bat and ball problem" is as follows: "A bat and a ball cost $1.10 in total. The bat costs $1.00 more than the ball. How much does the ball cost?" The intuitive but incorrect answer by most people here is 10 cents. The correct answer would be 5 cents. For the correct answer, it is necessary to override the intuitive answer by further reflection. In our study, 25% of the participants gave the correct answer and 67% gave the intuitive but incorrect answer (i.e., 10 cents). 8% of the participants gave other incorrect answers.

[5] Risk attitude was measured using a simple hypothetical gamble. Participants were asked to imagine that an ordinary coin was tossed and that they would receive zero euros if the coin came up heads and 100 euros if the coin came up tails. The participants were asked what they would be willing to pay (between 0 euros and 100 euros) to participate in the gamble. On average, our participants were very risk-averse with 12.86 euros.

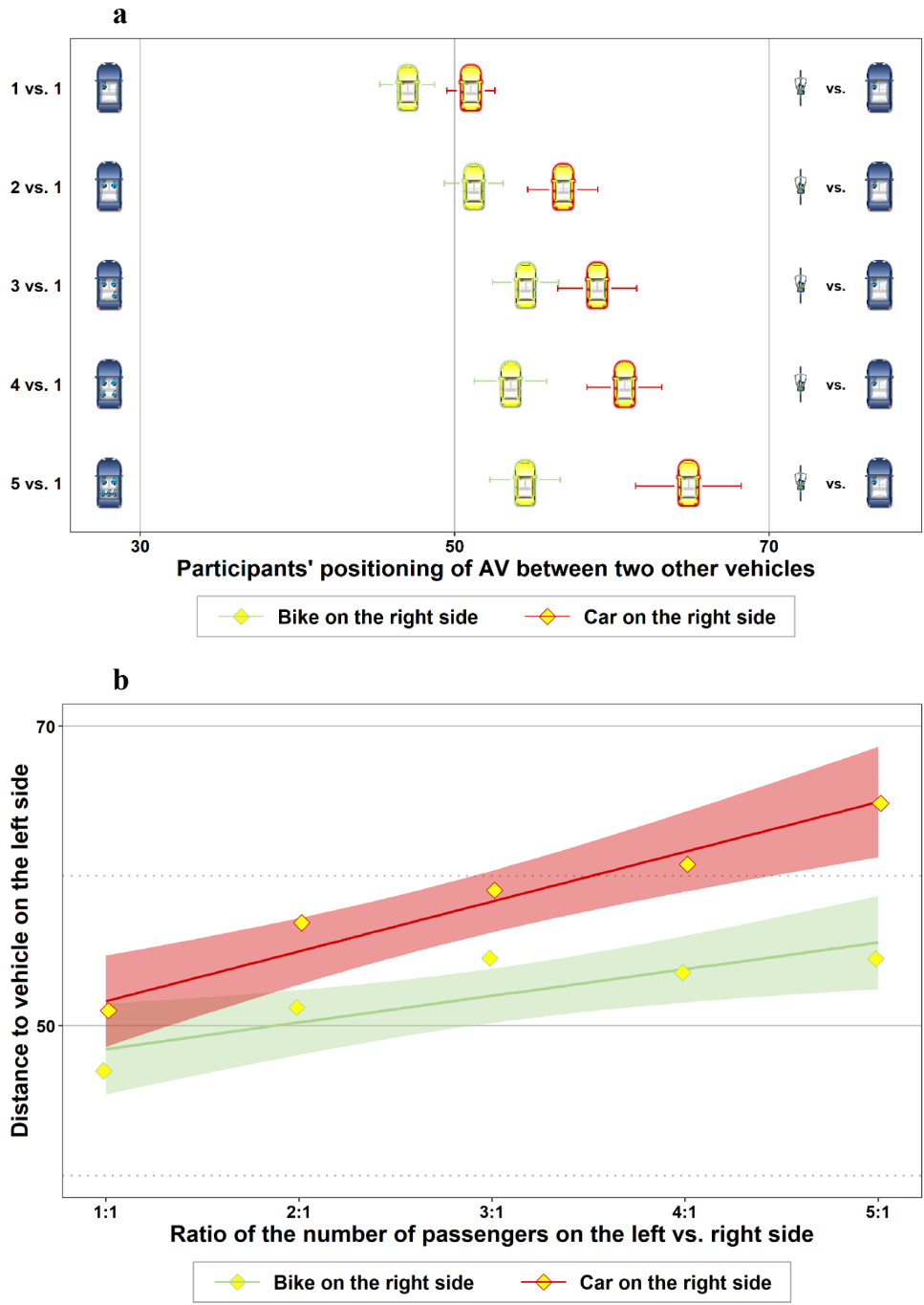

**Fig. 3. Risk allocation between different types of road users.**
(a) shows the means and standard errors of participants' chosen driving position of the AV in traffic situations with different numbers of passengers in the blue car on the left side and different road users on the right side. AVs with red frames depict the results when there is a car on the right side; AVs with green frames when there is a bicyclist on the right side. (b) shows the results of a linear regression in which the AV's distance from the car on the left side was regressed on the increasing disproportion of passengers on the left and right sides. We interacted the disproportion of passengers with a dummy variable for the two treatments (cyclist on the right yes/no). Red and green regions visualize 95% confidence intervals.

|  | Dependent variable: Distance to more passengers | | |
|---|---|---|---|
|  | AV empty | AV with passenger | Bike on the right side |
| Constant | 45.99*** (3.20) | 49.84*** (4.35) | 38.25*** (4.59) |
| Ratio of passengers | 2.33*** (0.59) | 2.03*** (0.57) | 1.77** (0.64) |
| More passengers on right side (= 1) | −3.63** (1.38) | −3.74* (1.89) |  |
| Decision time | −0.02 (0.04) | 0.00 (0.01) | 0.00 (0.03) |
| Initial position of AV | 0.03 (0.02) | 0.02 (0.03) | 0.03 (0.03) |
| Female (= 1) | 1.35 (1.40) | −2.35 (1.71) | 3.42 (2.00) |
| Age (in years) | 0.05 (0.04) | −0.00 (0.06) | 0.11 (0.07) |
| Job in technology sector (= 1) | 2.53 (1.77) | 3.79 (2.75) | 0.23 (3.06) |
| No driver license (= 1) | −0.37 (3.72) | 5.17 (3.34) | 2.63 (3.60) |
| Risk attitude | −0.01 (0.04) | −0.03 (0.05) | 0.02 (0.06) |
| "Bat and ball probl." (Correct = 1) | 3.09* (1.39) | −1.23 (1.62) | 0.08 (1.97) |
| Observations | 795 | 489 | 521 |
| Log Likelihood | −3445.24 | −2104.79 | −2322.33 |
| AIC | 6914.5 | 4233.6 | 4666.7 |

*p<0.05 **p<0.01 ***p<0.001

**Table 1: Coefficients (standard errors) of regressions for each treatment with the AVs' safety distance to more passengers as dependent variable.**

## Discussion

It is misguided that moral dilemmas in road traffic only occur in unavoidable accident scenarios that simply call for emergency braking (Cunneen et al., 2020; Davnall, 2020; Martinho et al., 2021). Participation in regular road traffic entails a distribution of risks between road users, which raises ethically relevant questions, especially when this distribution is enforced by AVs. The complexities of risk, however, pose a challenge to moral theory. One fruitful approach could be to view risk impositions as acceptable if they are supported by consensus (Hansson 2018). A driving algorithm whose training data is based on the value judgments of a representative sample of the population could be interpreted as a tool for building aggregate consensus. To reflect the moral intuitions of the German population, for instance, this algorithm would have to account for the number of potential victims and the type of road users in the AVs' driving behavior. This is something that might not happen if the intuition of mere accidence avoidance explicitly or implicitly determines the programming of driving algorithms.

Even if people are asked to imagine that they are in the passenger's role, they state their willingness to accept a higher accident probability for themselves if this decreases the probability of a more severe accident for other people. This is noteworthy with respect to the social dilemma of autonomous vehicles that has been identified in the context of unavoidable accidents (Bonnefon et al., 2016). The dilemma is that people approve of utilitarian AVs and like others to buy them but would themselves prefer AVs that protect them at all costs. We find that people express more altruism in the risky domain than they do in the deterministic domain. A person may be much more willing to rescue a child from a burning building if she attributes to herself a positive probability of survival. This divergence in stated preferences provides another reason to shift the ethics of AVs from the extreme case of unavoidable accidents to the regular case of mundane traffic maneuvers.